\documentclass[preprintnumbers,amsmath,amssymb,nofootinbib,superscriptaddress,floatfix,onecolumn,reprint,groupedaddress,eqsecnum,amsfonts,aps]{revtex4}

\pdfoutput=1
\usepackage[normalem]{ulem}%下划线宏包：\uline{my}可以在正文中my下面加下划线；\uuline{my} 可以在正文中my下面加双下划线；\uwave{my}下面加波浪线；\sout{my}将这个词用水平线删除；\xout{my}将这个词用斜线删除；
\usepackage{graphicx}
\usepackage[colorlinks,citecolor=blue,urlcolor=blue,linkcolor=blue]{hyperref}% 超链接宏包

\usepackage{subfigure}%关于图形的宏包，其说明文件已经和该论文放在一起
\usepackage{latexsym}
\usepackage{amssymb}
\usepackage{mathrsfs}
\usepackage{amsmath}
\usepackage{amsthm}
\usepackage{color}
\usepackage{dcolumn}% Align table columns on decimal point
\usepackage{bm}% bold math
\definecolor{dyellow}{rgb}{1.,0.8,.0}
\definecolor{myblue}{rgb}{.1,.1,.7}
\definecolor{dcyan}{rgb}{.0,.6,.6}
\definecolor{dmagenta}{rgb}{0.6,0.0,0.6}
\definecolor{brown}{rgb}{0.6,0.2,0.}
\definecolor{darkblue}{rgb}{.0,.0,0.5}
\definecolor{darkred}{rgb}{0.75,0.0,0.0}
\definecolor{orange}{rgb}{1.,.6,.0}
\definecolor{dorange}{rgb}{0.8,.4,.0}
\definecolor{darkgreen}{rgb}{0.0,0.6,0.0}
\definecolor{purple}{rgb}{.4,.0,.4}
\definecolor{grey}{rgb}{0.5,0.5,0.5}
\def\black{\color{black}}

\begin{document}
\hyphenpenalty=1000%等号后数值越大，可以更可能减少换行时断字太多的问题
%\preprint{APS/123-QED}
\title{The gravitational field outside a spatially compact stationary source in a generic fourth-order theory of gravity}

\author{Bofeng Wu}\email{bofengw@pku.edu.cn}%\affiliation{\NEU}
\affiliation{Department of Physics, College of Sciences, Northeastern University, Shenyang 110819, China}

\author{Chao-Guang Huang}%\footnote{Corresponding author}}
\email{huangcg@ihep.ac.cn}
\affiliation{Institute of High Energy Physics, Chinese Academy of Sciences, Beijing, 100049, China}

%\date{\today}
\begin{abstract}
By applying the symmetric and trace-free formalism in terms of the irreducible Cartesian tensors, the metric for the external gravitational field of a spatially compact stationary source is provided in $F(X,Y,Z)$ gravity, a generic fourth-order theory of gravity, \black where $X:=R$ is Ricci scalar, $Y:=R_{\mu\nu}R^{\mu\nu}$ is Ricci square, and $Z:=R_{\mu\nu\rho\sigma}R^{\mu\nu\rho\sigma}$ is Riemann square. A new type of gauge condition is proposed so that the linearized gravitational field equations of $F(X,Y,Z)$ gravity are greatly simplified, and then, the stationary metric in the region exterior to the source is derived. In the process of applying the result, integrations are performed only over the domain occupied by the source. The multipole expansion of the metric potential in $F(X,Y,Z)$ gravity for a spatially compact stationary source is also presented. In the expansion, the corrections of $F(X,Y,Z)$ gravity to General Relativity are Yukawa-like ones, dependent on two characteristic lengths. Two additional sets of mass-type source multipole moments appear in the corrections and the salient feature characterizing them is that the integrations in their expressions are always modulated by a common radial factor related to the source distribution.
\end{abstract}
%\pacs{04.50.Kd,  04.25.-g, 04.25.Nx}
\maketitle
\section{Introduction}
Although General Relativity (GR) survives in many tests~\cite{TheLIGOScientific:2016agk,TheLIGOScientific:2016agk1,Clifford2018},
it still faces many challenges, and one of the typical examples is that GR can not give a widely accepted explanation to the cosmic acceleration without introduction of dark energy~\cite{cosmicacceleration}. Introduction of the alternative theories of gravity is one approach to handling these difficulties~\cite{Sotiriou2010,Clifford2018,Capozziello:2011et}, and thus, identifying the correct theory of gravity is a crucial issue of modern physics.
In this work, we shall focus our attention on the fourth-order theories of gravity in the metric formalism, which modifies the Einstein-Hilbert action by adopting a general function of curvature invariants in the gravitational Lagrangian. Since derivatives of curvature invariants like $\square R$ and the parity-odd Chern-Simons invariant~\cite{Alexander:2009tp} that enters at the same order in curvatures and derivatives are not considered, the gravitational field equations of such theories are fourth-order in derivatives of the metric tensor~\cite{Clifton:2011jh,stabile2015}.

$f(R)$ gravity~\cite{Gibbons1977,Starobinsky:1980te,Nojiri:2010wj,DNojiri:2017ncd} is one of the simplest fourth-order theories of gravity. Besides the Ricci scalar $R$, there are several other curvature invariants that one can construct from the metric, and all of them are the combinations of contractions of the Riemann tensor one or more times with itself and the metric~\cite{Thomas2007}. $F(X,Y,Z)$ gravity, being a generic fourth-order theory of gravity, is obtained by replacing the Einstein-Hilbert action in the gravitational Lagrangian by a general function $F$ of curvature invariants $X,Y,$ and $Z$~\cite{stabile2015,Stabile:2010mz,Capozziello:2009ss,Bogdanos:2009tn}, where $X:=R,\ Y:=R_{\mu\nu}R^{\mu\nu},$ and $Z:=R_{\mu\nu\rho\sigma}R^{\mu\nu\rho\sigma}$ with $R_{\mu\nu}$ as the Ricci tensor and $R_{\mu\nu\rho\sigma}$ as the Riemann tensor. $F(X,Y,Z)$ gravity contains a large number of sub-models, such as GR, Starobinsky gravity, $f(R)$ gravity, and $f(R,\mathcal{G})$ gravity ($\mathcal{G}$ is the Gauss-Bonnet scalar), etc., and they all have a wide range of applications in
gravitational physics~\cite{Gibbons1977,Starobinsky:1980te,Nojiri:2010wj,DNojiri:2017ncd,Cognola:2006eg,Alimohammadi:2008fq,Bamba:2009uf,Wu:2015maa,Shamir:2017ndy,Odintsov:2018nch}.
%It could be expected that if a certain result of some specific physical phenomenon is given in $F(X,Y,Z)$ gravity, the corresponding one in its any sub-model can also be derived. As a result, $F(X,Y,Z)$ gravity actually provides a general theoretical framework for fourth-order theories of gravity.

In order to test $F(X,Y,Z)$ gravity, the metric for the external gravitational field of a gravitating source should be deduced. However, since the gravitational field equations of $F(X,Y,Z)$ gravity are exceedingly complicated, one usually has to adopt some approximation method to simplify them. In Refs.~\cite{stabile2015,Capozziello:2009ss,Stabile:2010mz}, these equations are handled under the weak-field and slow-motion approximation, and then, with the Green function method, the expressions of the metric for the gravitational field generated by a source are provided. Although these results are valid in general, they seem to be inconvenient to apply
%additional integration techniques are required when one tries to apply them
in practice because integrations involved in these expressions usually need to be carried out over the whole space.
In this paper, we will make an attempt to redeal with this problem under the weak-field approximation, and the key point is how to simplify the linearized gravitational field equations of $F(X,Y,Z)$ gravity.
%Even so, these equations are still difficult to be disposed of, which results in that the external metric for a spatially compact stationary source has not been obtained yet.
%We will make an attempt to solve this problem in this paper, and the key point is how to simplify the linearized gravitational field equations of $F(X,Y,Z)$ gravity.
Motivated by the method in Refs.~\cite{Wu:2017vvm,Wu:2018hjx,Wu:2018jve,Wu:2021uws}, a new type of gauge condition is proposed so that these linearized equations can be successfully transformed into d'Alembert equation and Klein-Gordon equations with external sources. Then, with the help of the symmetric and trace-free (STF) formalism in terms of the irreducible Cartesian tensors, developed by Thorne~\cite{Thorne:1980ru}, Blanchet, Damour, and Iyer~\cite{Blanchet:1985sp,Blanchet:1989ki,Damour:1990gj}, the stationary metric in the region exterior to the source is presented.
Compared with those in Refs.~\cite{stabile2015,Capozziello:2009ss,Stabile:2010mz}, the salient feature of the result in the present paper is that the integrations are performed just over the region occupied by the source, which could lead to a wider application of the result in this paper.

The metric consists of the GR-like part and the modified part, where the former is exactly the result in GR when $F(X,Y,Z)$ gravity reduces to GR, and the latter is the correction to the former in $F(X,Y,Z)$ gravity. As with previous results in Refs.~\cite{stabile2015,Capozziello:2009ss,Stabile:2010mz}, the modified part of the metric is also characterized by two characteristic lengths depending on the value of derivatives of $F$ with respect to curvature invariants, which implies that in $F(X,Y,Z)$ gravity, there are two massive propagations in general. In addition, it should be noted that the multipole expansion of the metric potential in $F(X,Y,Z)$ gravity for a spatially compact stationary source is also presented in the present paper. In this expansion, the GR-like part, identical to the metric potential in GR, is associated with the mass multipole moments, and the modified part, representing the Yukawa-like corrections to the GR-like part, is associated with two additional sets of mass-type source multipole moments. In the expressions of these two sets of source multipole moments, the integrations are always modulated by a common radial factor related to the source, which implies that differently from the case in GR, the effects even at the monopole order in $F(X,Y,Z)$ gravity also depend on the source distribution.
%Moreover, it should be noted that since the metric obtained in the present paper is applicable for the external gravitational field of any spatially compact stationary source, it is suitable to be used to explore phenomena happening in the gravitational field outside a realistic gravitating source because the effect of the size and shape of the source is extremely crucial. As a consequence, it is certain that our result will make $F(X,Y,Z)$ gravity have a wider range of applications.

As mentioned before, a large number of fourth-order theories of gravity are sub-models of $F(X,Y,Z)$ gravity, and therefore, the metrics for the external gravitational field of a spatially compact stationary source in these models can be directly obtained from the one in $F(X,Y,Z)$ gravity under certain conditions. As examples, the corresponding metrics in GR, $f(R)$ gravity, and $f(R,\mathcal{G})$ gravity are all presented in this paper. It is shown that when $F(X,Y,Z)\rightarrow f(R)$ or $f(R,\mathcal{G})$, one characteristic length of the metric in $F(X,Y,Z)$ gravity disappears, and the metric in $f(R,\mathcal{G})$ gravity is the same as that in $f(R)$ gravity, which is consistent with the fact that the Gauss-Bonnet scalar $\mathcal{G}$, being a topological invariant, has no contribution to the gravitational field dynamics. Furthermore, when $F(X,Y,Z)$ gravity reduces to GR, both the characteristic lengths disappear, and the metric recovers the classical one in GR.

It is very meaningful to explore the effects of the additional gravitational degrees of freedom appearing in $F(X,Y,Z)$ gravity, where these new degrees of freedom could be analyzed from the modified part of the metric. To this end, the metric needs to be applied to some specific phenomenon in practice. For a gyroscope moving around the source in geodesic motion, one is able to use the metric to derive its spin's angular velocity of precession, and according to the conventional method in Ref.~\cite{MTW1973}, the precessional angular velocity in GR would be corrected by the modified part of the metric. Then, by comparing these results with the data of the gyroscopic experiment, e.g., Gravity Probe B (GP-B), the effects of the new gravitational degrees of freedom in $F(X,Y,Z)$ gravity will be obtained. Similarly, the metric can also be applied to the anomalous perihelion advance of Mercury, the gravitational redshift of light, and the light bending, etc. By comparing the theoretical results with the experimental or observational data, the further effects of the new gravitational degrees of freedom in $F(X,Y,Z)$ gravity will also be derived.
\black
%The increment or reduction of the characteristic lengths in different sub-models of $F(X,Y,Z)$ gravity implies that the theoretical predictions of these models for a physical phenomenon are completely different. Hence,
%if the metrics in these sub-models of $F(X,Y,Z)$ gravity are applied to dealing with some specific phenomenon, by comparing the resulting theoretical results with the experimental or observational data, the constraints on the coefficients of curvature invariants in the gravitational Lagrangians of these models may be obtained, and as a consequence, by such approach, a large class of fourth-order theories of gravity could be assessed in detail.

This paper is organized as follows. In Sec.~\ref{Sec:second}, the notation and relevant formulas in the STF formalism are briefly reviewed.
In Sec.~\ref{Sec:third}, a weak-field approximation method is developed within $F(X,Y,Z)$ gravity. In Sec.~\ref{Sec:fourth}, the metric for the external gravitational field of a spatially compact stationary source in $F(X,Y,Z)$ gravity is derived. In Sec.~\ref{Sec:fifth}, the conclusions and the related discussions are presented.
%%%%%%%%%%%%%%%%%%%%%%%%%%%%%%%%%%%%%%%%%%%%%%%%%%%%%%%
%%%%%%%%%%%%%%%%%%%%%%%%%%%%%%%%%%%%%%%%%%%%%%%%%%%%%%%
%%% The first section                              %%%%
%%%%%%%%%%%%%%%%%%%%%%%%%%%%%%%%%%%%%%%%%%%%%%%%%%%%%%%
%%%%%%%%%%%%%%%%%%%%%%%%%%%%%%%%%%%%%%%%%%%%%%%%%%%%%%%
\section{Notation and relevant formulas in the STF formalism~\cite{Wu:2017vvm,Wu:2018hjx,Wu:2018jve,Wu:2021uws,Damour:1990gj}\label{Sec:second}}
Throughout the paper, the following notation and rules are adopted:
\begin{itemize}
\item The international system of units is employed;
\item The Greek letters, representing the spacetime indices, range from 0 to 3, and the Latin letters, representing the space indices, range from 1 to 3;
\item Repeated indices appearing within a term indicate that the sum should be taken over;
\item The signature of the spacetime metric $g_{\mu\nu}$ is $(-,+,+,+)$;
\item In the linearized gravity theory, the coordinates $(x^{\mu})=(ct,x_{i})$ are treated as the Minkowskian coordinates;
\item The spherical coordinate system $(ct,r,\theta,\varphi)$ is given by
\begin{equation}\label{equ2.1}
x_{1}=r\sin{\theta}\cos{\varphi},\ x_{2}=r\sin{\theta}\sin{\varphi},\ x_{3}=r\cos{\theta};
\end{equation}
\item The radial vector in the flat space is $\boldsymbol{x}=x_{i}\partial_{i}$ with $x_{i}$ as the components and $\partial_{i}:=\partial/\partial x_{i}$ as the coordinate basis vectors. Let $r=|\boldsymbol{x}|$ be the length of $\boldsymbol{x}$, and then, $\boldsymbol{n}=\boldsymbol{x}/r=n_{i}\partial_{i}$ is the unit radial vector with $n_{i}=x_{i}/r$.
%Obviously, the following equality holds.
%\begin{equation}\label{equ2.2}
%\partial_{r}:=\frac{\partial}{\partial r}=n_{i}\partial_{i}=\boldsymbol{n}.
%\end{equation}
\end{itemize}
The STF part of a Cartesian tensor $B_{I_{l}}:=B_{i_{1}i_{2}\cdots i_{l}}$ is defined by
\begin{eqnarray}
\label{equ2.2}
\hat{B}_{I_{l}}:=B_{\langle I_{l}\rangle}=B_{\langle i_{1}i_{2}\cdots i_{l}\rangle}
:=\sum_{k=0}^{\left[\frac{l}{2}\right]}b_{k}\delta_{(i_{1}i_{2}}\cdots\delta_{i_{2k-1}i_{2k}}S_{i_{2k+1}\cdots i_{l})a_{1}a_{1}\cdots a_{k}a_{k}},
\end{eqnarray}
where
\begin{equation}
\label{equ2.3}b_{k}:=(-1)^{k}\frac{(2l-2k-1)!!}{(2l-1)!!}\frac{l!}{(2k)!!(l-2k)!},
\end{equation}
and
\begin{equation}\label{equ2.4}
S_{I_{l}}:=B_{(I_{l})}=B_{(i_{1}i_{2}\cdots i_{l})}:=\frac{1}{l!}\sum_{\sigma} B_{i_{\sigma(1)}i_{\sigma(2)}\cdots i_{\sigma(l)}}
\end{equation}
is its symmetric part with $\sigma$ running over all permutations of $(12\cdots l)$. The quantities
\begin{eqnarray}
\label{equ2.5}X_{I_{l}}&=&X_{i_{1}i_{2}\cdots i_{l}}:= x_{i_{1}}x_{i_{2}}\cdots x_{i_{l}},\\
\label{equ2.6}N_{I_{l}}&=&N_{i_{1}i_{2}\cdots i_{l}}:= n_{i_{1}}n_{i_{2}}\cdots n_{i_{l}}
\end{eqnarray}
are used to denote the tensor products of $l$ radial and unit radial vectors, respectively, and they satisfy
\begin{equation}\label{equ2.7}
X_{I_{l}}=r^l N_{I_{l}}.
\end{equation}
Other relevant formulas in the STF formalism include:
\begin{eqnarray}
\label{equ2.8}\hat{N}_{I_{l}}&=&\sum_{k=0}^{\left[\frac{l}{2}\right]}b_{k}\delta_{(i_{1}i_{2}}\cdots\delta_{i_{2k-1}i_{2k}}
N_{i_{2k+1}\cdots i_{l})},\\
\label{equ2.9}\hat{\partial}_{I_{l}}&=&\sum_{k=0}^{\left[\frac{l}{2}\right]}b_{k}\delta_{(i_{1}i_{2}}\cdots\delta_{i_{2k-1}i_{2k}}
\partial_{i_{2k+1}\cdots i_{l})}\left(\nabla^2\right)^k,\\
\label{equ2.10}\hat{\partial}_{I_{l}}\left(\frac{F(r)}{r}\right)&=&\hat{N}_{I_{l}}\sum_{k=0}^{l}\frac{(l+k)!}{(-2)^{k}k!(l-k)!}
\frac{\partial_{r}^{l-k}F(r)}{r^{k+1}},
\end{eqnarray}
where $\nabla^2=\partial_{a}\partial_{a}$ is the Laplace operator in flat space, $\partial_{I_{l}}=\partial_{i_{1}i_{2}\cdots i_{l}}:=\partial_{i_{1}}\partial_{i_{2}}\cdots\partial_{i_{l}}$,
and $\partial_r^{l-k}$ is the $(l-k)$-th derivative with respect to $r$.
%%%%%%%%%%%%%%%%%%%%%%%%%%%%%%%%%%%%%%%%%%%%%%%%%%%%%%%
%%%%%%%%%%%%%%%%%%%%%%%%%%%%%%%%%%%%%%%%%%%%%%%%%%%%%%%
%%% The second section                             %%%%
%%%%%%%%%%%%%%%%%%%%%%%%%%%%%%%%%%%%%%%%%%%%%%%%%%%%%%%
%%%%%%%%%%%%%%%%%%%%%%%%%%%%%%%%%%%%%%%%%%%%%%%%%%%%%%%
\section{The Weak-field approximation of $F(X,Y,Z)$ gravity~\label{Sec:third}}
The action of $F(X,Y,Z)$ gravity~\cite{stabile2015,Capozziello:2009ss,Stabile:2010mz} is
\begin{equation}
\label{equ3.1}S=\frac{1}{2\kappa c}\int d^4x\sqrt{-g}F(X,Y,Z)+S_{M}(g^{\mu\nu},\psi),
\end{equation}
where $\kappa=8\pi G/c^{4}$ with $G$ as the gravitational constant, $c$ is the velocity of light in vacuum, $g$ is the determinant of metric $g_{\mu\nu}$, and $S_{M}(g^{\mu\nu},\psi)$ is the matter action. In the metric formalism, the gravitational field equations and the corresponding trace equation of $F(X,Y,Z)$ gravity are given by varying the above action with respect to $g^{\mu\nu}$,
\begin{equation}
\label{equ3.2}H_{\mu\nu}=\kappa T_{\mu\nu},\qquad H=\kappa T,
\end{equation}
where
\begin{align}
\label{equ3.3}H_{\mu\nu}=&-\frac{1}{2}g_{\mu\nu}F+(R_{\mu\nu}+g_{\mu\nu}\square -\nabla_{\mu}\nabla_{\nu}) F_{X}-2\nabla^{\lambda}\nabla_{(\mu} (F_{Y}R_{\nu)\lambda})+g_{\mu\nu}\nabla^{\alpha}\nabla^{\beta}(F_{Y}R_{\alpha\beta})\notag\\
&+\square(F_{Y}R_{\mu\nu})+2F_{Y}R_{\mu\alpha}R^{\alpha}_{\phantom{\alpha}\nu}+
4\nabla^{(\rho}\nabla^{\sigma)}(F_{Z}R_{\mu\rho\nu\sigma})+2F_{Z}R_{\mu}^{\phantom{\mu}\alpha\beta\gamma}
R_{\nu\alpha\beta\gamma},\\
\label{equ3.4}H=&-2F+F_{X}X+2F_{Y}R_{\mu\nu}R^{\mu\nu}+2F_{Z}R_{\nu\alpha\beta\gamma}R^{\nu\alpha\beta\gamma}+3\square F_{X}+\square (F_{Y}X)\notag\\
&+2\nabla^{\alpha}\nabla^{\beta}(F_{Y}R_{\alpha\beta})+4\nabla^{\rho}\nabla^{\sigma}(F_{Z}R_{\rho\sigma})
\end{align}
with $F_{X}:=\partial F/\partial X$, $F_{Y}:=\partial F/\partial Y$, and $F_{Z}:=\partial F/\partial Z$, and $T_{\mu\nu}$
is the energy-momentum tensor of matter with $T=g_{\mu\nu}T^{\mu\nu}$ as its trace. As in Refs.~\cite{stabile2015,Stabile:2010mz}, we assume that $F(X,Y,Z)$ is expressed as a power series
\begin{equation}
\label{equ3.5}F(X,Y,Z)=X+F_{2}Y+F_{3}Z+\frac{1}{2}\left(F_{11}X^2+F_{22}Y^2+F_{33}Z^2+2F_{12}XY+2F_{13}XZ+2F_{23}YZ\right)+\cdots,
\end{equation}
where the dimensions of the coefficients $F_{2},F_{3},F_{11},F_{22}\cdots$ are $[X]^{-1},[X]^{-1},[X]^{-1},[X]^{-3}\cdots$, respectively. Such models are physically interesting and allow to recover the results of GR and the correct boundary and asymptotic conditions in general.

Equations~(\ref{equ3.2})---(\ref{equ3.4}) show that the gravitational field equations of $F(X,Y,Z)$ gravity are exceedingly complicated, we will first adopt the weak-field approximation to simplify them. Let $\eta^{\mu\nu}$ be the Minkowskian metric in a fictitious flat spacetime, and the gravitational field amplitude $h^{\mu\nu}$ is defined as
\begin{eqnarray}
\label{equ3.6}h^{\mu\nu}&:=&\sqrt{-g}g^{\mu\nu}-\eta^{\mu\nu}.
\end{eqnarray}
Just as in GR~\cite{Blanchet:2013haa,eric2018}, $h^{\mu\nu}$, being a perturbation in the linearized framework of the weak-field approximation, satisfies
\begin{equation}
\label{equ3.7}|h^{\mu\nu}|\ll1,
\end{equation}
and then, the linear parts (denoted by the superscript (1)) of the Riemann tensor, the Ricci tensor and the Ricci scalar
are, respectively,
\begin{align}
\label{equ3.8}R^{\mu\nu\rho\sigma(1)}=&\ -\frac{1}{2}
\big(\partial^{\rho}\partial^{\nu}h^{\mu\sigma}
-\partial^{\sigma}\partial^{\nu}h^{\mu\rho}
+\partial^{\sigma}\partial^{\mu}h^{\nu\rho}
-\partial^{\rho}\partial^{\mu}h^{\nu\sigma}\big)\notag\\
&\ -\frac{1}{4}\big(\eta^{\mu\rho}\partial^{\sigma}\partial^{\nu}h
-\eta^{\mu\sigma}\partial^{\rho}\partial^{\nu}h+\eta^{\nu\sigma}\partial^{\rho}\partial^{\mu}h
-\eta^{\nu\rho}\partial^{\sigma}\partial^{\mu}h\big),\\
\label{equ3.9}R^{\mu\nu(1)}=&\ \frac{1}{2}\square_{\eta}h^{\mu\nu}-\frac{1}{4}\eta^{\mu\nu}\square_{\eta}h-\frac{1}{2}\big(\partial_{\lambda}\partial^{\mu}h^{\nu\lambda}+\partial_{\lambda}\partial^{\nu}h^{\mu\lambda}\big),\\
\label{equ3.10}X^{(1)}=&\ R^{(1)}=-\frac{1}{2}\square_{\eta}h-\partial_{\alpha}\partial_{\beta}h^{\alpha\beta}
\end{align}
with $\partial^{\rho}:=\eta^{\rho\sigma}\partial_{\sigma}=\eta^{\rho\sigma}\partial/\partial x^{\sigma}$, $h=\eta_{\mu\nu}h^{\mu\nu}$, and $\square_{\eta}:=\eta^{\mu\nu}\partial_{\mu}\partial_{\nu}$. Thus, from Eq.~(\ref{equ3.5}), $F,F_{X},F_{Y},$ and $F_{Z}$ can be written as
\begin{eqnarray}
\label{equ3.11}&&\left\{\begin{array}{l}
\displaystyle F=X^{(1)}+o(h^{\mu\nu}),\smallskip\\
\displaystyle F_{X}=1+F_{11}X^{(1)}+o(h^{\mu\nu}),\smallskip\\
\displaystyle F_{Y}=F_{2}+F_{12}X^{(1)}+o(h^{\mu\nu}),\smallskip\\
\displaystyle F_{Z}=F_{3}+F_{13}X^{(1)}+o(h^{\mu\nu}),
\end{array}\right.
\end{eqnarray}
where $o(h^{\mu\nu})$ is the higher order terms of $h^{\mu\nu}$. By virtue of Eqs.~(\ref{equ3.6})---(\ref{equ3.11}), the linearized gravitational field equations and the corresponding trace equation of $F(X,Y,Z)$ gravity can be derived from Eqs.~(\ref{equ3.2})---(\ref{equ3.4}),
\begin{eqnarray}
\label{equ3.12}H^{\mu\nu(1)}&=&R^{\mu\nu(1)}+\big(F_{2}+4F_{3}\big)\square_{\eta}R^{\mu\nu(1)}-\frac{1}{2}\eta^{\mu\nu}X^{(1)}+\left(F_{11}+\frac{F_{2}}{2}\right)\eta^{\mu\nu}\square_{\eta}X^{(1)}\notag\\
&&-\big(F_{11}+F_{2}+2F_{3}\big)\partial^{\mu}\partial^{\nu}X^{(1)}=\kappa T^{\mu\nu},\\
\label{equ3.13}H^{(1)}&=&-X^{(1)}+\big(3F_{11}+2F_{2}+2F_{3}\big)\square_{\eta}X^{(1)}=\kappa T.
\end{eqnarray}
It should be noted that $T^{\mu\nu}$ and $T$ in the above two equations are the energy-momentum tensor of matter living in Minkowski spacetime and its trace, respectively. One may observe that these linearized equations are still difficult to be disposed of, which implies that we need to find a new method to further simplify them. A basic fact is that since the Lagrangian of $F(X,Y,Z)$ gravity is a function of the gauge-invariant curvature scalars, its gravitational field equations are gauge-invariant, and as a result, we have the freedom to perform a gauge transformation~\cite{Berry:2011pb}. Motivated by the approach in Refs.~\cite{Berry:2011pb,Wu:2017vvm,Wu:2018hjx,Wu:2018jve,Wu:2021uws},
we construct the effective gravitational field amplitude of the linearized $F(X,Y,Z)$ gravity as
\begin{eqnarray}
\label{equ3.14}\tilde{h}^{\mu\nu}&:=&h^{\mu\nu}+a\eta^{\mu\nu}X^{(1)}+bR^{\mu\nu(1)},
\end{eqnarray}
where $a$ and $b$ are the parameters to be specified. In this paper, the gauge condition
\begin{equation}\label{equ3.15}
\partial_{\mu}\tilde{h}^{\mu\nu}=0
\end{equation}
will be imposed, and one will see that it is due to this condition that the linearized gravitational field equations of $F(X,Y,Z)$ gravity can be transformed into d'Alembert equation and Klein-Gordon equations with external sources. Firstly, by inserting Eqs.~(\ref{equ3.14}) and (\ref{equ3.15}) into Eqs.~(\ref{equ3.9}) and (\ref{equ3.10}), $R^{\mu\nu(1)}$ and $X^{(1)}$ are reexpressed as
\begin{align}
\label{equ3.16}R^{\mu\nu(1)}&=\frac{1}{2}\square_{\eta}\tilde{h}^{\mu\nu}-\frac{1}{4}\eta^{\mu\nu}\square_{\eta}\tilde{h}+\frac{2a+b}{4}\eta^{\mu\nu}\square_{\eta}X^{(1)}-\frac{b}{2}\square_{\eta}R^{\mu\nu(1)}+\left(a+\frac{b}{2}\right)\partial^{\mu}\partial^{\nu}X^{(1)},\\
\label{equ3.17}X^{(1)}&=-\frac{1}{2}\square_{\eta}\tilde{h}+(3a+b)\square_{\eta}X^{(1)},
\end{align}
and then, substituting them in Eqs.~(\ref{equ3.12}) and (\ref{equ3.13}), one gets
\begin{eqnarray}
\label{equ3.18}H^{\mu\nu(1)}&=&\frac{1}{2}\square_{\eta}\tilde{h}^{\mu\nu}+\left(F_{11}+\frac{F_{2}}{2}-a-\frac{b}{4}\right)\eta^{\mu\nu}\square_{\eta}X^{(1)}+\left(F_{2}+4F_{3}-\frac{b}{2}\right)\square_{\eta}R^{\mu\nu(1)}\notag\\
&&-\left(F_{11}+F_{2}+2F_{3}-a-\frac{b}{2}\right)\partial^{\mu}\partial^{\nu}X^{(1)}=\kappa T^{\mu\nu},\\
\label{equ3.19}H^{(1)}&=&\frac{1}{2}\square_{\eta}\tilde{h}+\big(3F_{11}+2F_{2}+2F_{3}-3a-b\big)\square_{\eta}X^{(1)}=\kappa T.
\end{eqnarray}
If we pick
\begin{eqnarray}
\label{equ3.20}&&\left\{\begin{array}{l}
\displaystyle a=F_{11}-2F_{3},\smallskip\\
\displaystyle b=2F_{2}+8F_{3},
\end{array}\right.
\end{eqnarray}
Eqs.~(\ref{equ3.18}) and (\ref{equ3.19}) reduce to
\begin{eqnarray}
\label{equ3.21}\square_{\eta}\tilde{h}^{\mu\nu}&=&2\kappa T^{\mu\nu},\\
\label{equ3.22}\square_{\eta}\tilde{h}&=&2\kappa T.
\end{eqnarray}
Obviously, the effective gravitational field amplitude $\tilde{h}^{\mu\nu}$ satisfies d'Alembert equation, and therefore, it should behave just as its counterpart in GR.
The physical meaning of the gauge condition~(\ref{equ3.15}) can be found from Eq.~(\ref{equ3.20}). By plugging Eq.~(\ref{equ3.20}) into Eq.~(\ref{equ3.14}) and then making use of the linearized Bianchi's identity $\partial_{\mu}R^{\mu\nu(1)}=\partial^{\nu}X^{(1)}/2$,
one can acquire
\begin{eqnarray}
\label{equ3.23}\partial_{\mu}h^{\mu\nu}=-\big(F_{11}+F_{2}+2F_{3}\big)\partial^{\nu}X^{(1)},
\end{eqnarray}
which explicitly shows that the gauge condition~(\ref{equ3.15}) is no longer the harmonic gauge condition $\partial_{\mu}h^{\mu\nu}=0$~\cite{Clifford2018}.
Now, if we substitute Eq.~(\ref{equ3.20}) back in Eqs.~(\ref{equ3.16}) and (\ref{equ3.17}) and introduce the quantities
\begin{eqnarray}
\label{equ3.24}m_{1}^2&:=&\frac{1}{3F_{11}+2F_{2}+2F_{3}},\\
\label{equ3.25}m_{2}^2&:=&-\frac{1}{F_{2}+4F_{3}},
\end{eqnarray}
the differential equations fulfilled by $R^{\mu\nu(1)}$ and $X^{(1)}$ are obtained,
\begin{eqnarray}
\label{equ3.26}R^{\mu\nu(1)}&=&\kappa\left(T^{\mu\nu}-\frac{1}{2}\eta^{\mu\nu}T\right)+\frac{1}{m_{2}^2}\square_{\eta}R^{\mu\nu(1)}+\frac{1}{3}\left(\frac{1}{m_{1}^2}-\frac{1}{m_{2}^2}\right)\left(\frac{1}{2}\eta^{\mu\nu}\square_{\eta}X^{(1)}+\partial^{\mu}\partial^{\nu}X^{(1)}\right),\\
\label{equ3.27}X^{(1)}&=&-\kappa T+\frac{1}{m_{1}^2}\square_{\eta}X^{(1)}.
\end{eqnarray}
Equation~(\ref{equ3.27}) is actually Klein-Gordon equation with an external source, namely
\begin{eqnarray}
\label{equ3.28}\square_{\eta}X^{(1)}-m_{1}^2 X^{(1)}&=&m_{1}^2\kappa T,
\end{eqnarray}
and in order to get a physically meaningful solution~\cite{Olmo:2005zr,Corda:2007nr}, we constrain $F(X,Y,Z)$ such that $m_{1}^2>0$. Thus,
as implied in Ref.~\cite{Stabile:2010mz}, it is clearly shown that the Ricci scalar $X^{(1)}$ presents a massive propagation.
Equation~(\ref{equ3.26}) is very complicated so that seeking its solution is a challenging task.
%, which is the key reason why the metric for the external gravitational field of a realistic source has not been obtained yet.
With the help of Eq.~(\ref{equ3.28}),  Eq.~(\ref{equ3.26}) can be greatly simplified. From Eq.~(\ref{equ3.28}), $\eta^{\mu\nu}\square_{\eta}X^{(1)}$ and $\partial^{\mu}\partial^{\nu}X^{(1)}$ in Eq.~(\ref{equ3.26}) have the following decompositions:
\begin{eqnarray}
\label{equ3.29}&&\left\{\begin{array}{l}
\displaystyle \eta^{\mu\nu}\square_{\eta}X^{(1)}=d_{1}\eta^{\mu\nu}m_{1}^2 X^{(1)}+(1-d_{1})\eta^{\mu\nu}\square_{\eta}X^{(1)}+d_{1}\eta^{\mu\nu}m_{1}^2\kappa T,\smallskip\\
\displaystyle \partial^{\mu}\partial^{\nu}X^{(1)}=\frac{d_{2}}{m_{1}^2}\partial^{\mu}\partial^{\nu}\big(\square_{\eta}X^{(1)}\big)+(1-d_{2})\partial^{\mu}\partial^{\nu}X^{(1)}-d_{2}\partial^{\mu}\partial^{\nu}(\kappa T),
\end{array}\right.
\end{eqnarray}
where $d_{1}$ and $d_{2}$ are two arbitrary parameters, and by applying (\ref{equ3.29}),
one is able to successfully rewrite Eq.~(\ref{equ3.26}) as
\begin{eqnarray}
\label{equ3.30}\square_{\eta}P^{\mu\nu}-m_{2}^2 P^{\mu\nu}&=&-m_{2}^2\kappa S^{\mu\nu}
\end{eqnarray}
with
\begin{eqnarray}
\label{equ3.31}P^{\mu\nu}&=&R^{\mu\nu(1)}-\frac{1}{6}\eta^{\mu\nu}X^{(1)}-\frac{1}{3m_{1}^2}\partial^{\mu}\partial^{\nu}X^{(1)},\\
\label{equ3.32}S^{\mu\nu}&=&T^{\mu\nu}-\frac{1}{3}\eta^{\mu\nu}T+\frac{1}{3m_{2}^2}\partial^{\mu}\partial^{\nu}T.
\end{eqnarray}
Equation~(\ref{equ3.30}) explicitly indicates that $P^{\mu\nu}$ also satisfies Klein-Gordon equation with an external source, and if $m_{2}^2>0$, the components of the tensor $P^{\mu\nu}$ also present massive propagations in $F(X,Y,Z)$ gravity. As in Refs.~\cite{Stabile:2010mz,stabile2015}, we will always choose $m_{1}^2>0$ and $m_{2}^2>0$ in this paper so as to obtain physically meaningful solutions to Eqs.~(\ref{equ3.28}) and (\ref{equ3.30}).
From the above process, we observe that by imposing the gauge condition (\ref{equ3.15}), the original linearized gravitational field equations (\ref{equ3.12}) of $F(X,Y,Z)$ gravity have indeed been converted to d'Alembert equation (\ref{equ3.21}) and Klein-Gordon equations (\ref{equ3.28}) and (\ref{equ3.30}). Since these equations are easy to handle, the following tasks are straightforward.

\section{Metric for the external gravitational field of a spatially compact stationary source in $F(X,Y,Z)$ gravity~\label{Sec:fourth}}
Next, we shall seek to find the solutions to Eqs.~(\ref{equ3.21}), (\ref{equ3.28}), and (\ref{equ3.30}) for a spatially compact stationary source with the help of the STF formalism presented in Sec.~\ref{Sec:second}. Since the source is time-independent, Eqs.~(\ref{equ3.21}), (\ref{equ3.28}), and (\ref{equ3.30}) reduce to
\begin{eqnarray}
\label{equ3.33}\nabla^{2}\tilde{h}^{\mu\nu}&=&2\kappa T^{\mu\nu},\\
\label{equ3.34}\nabla^{2}X^{(1)}-m_{1}^2 X^{(1)}&=&m_{1}^2\kappa T,\\
\label{equ3.35}\nabla^{2}P^{\mu\nu}-m_{2}^2 P^{\mu\nu}&=&-m_{2}^2\kappa S^{\mu\nu}.
\end{eqnarray}
According to relevant results in Ref.~\cite{Wu:2017vvm}, the solution to Eq.~(\ref{equ3.33}) for a spatially compact source is
\begin{equation}\label{equ3.36}
\left\{\begin{array}{l}
\displaystyle\tilde{h}^{00}(\boldsymbol{x})=\displaystyle -\frac{4G}{c^{2}}\sum_{l=0}^{\infty}\frac{(-1)^{l}}{l!}\hat{M}_{I_{l}}\partial_{I_{l}}\left(\frac{1}{r}\right),\\
\displaystyle\tilde{h}^{0i}(\boldsymbol{x})=\displaystyle-\frac{4G}{c^{3}}\sum_{l=1}^{\infty}\frac{(-1)^{l}l}{(l+1)!}\epsilon_{iab}\hat{S}_{aI_{l-1}}\partial_{bI_{l-1}}\left(\frac{1}{r}\right),\\
\displaystyle\tilde{h}^{ij}(\boldsymbol{x})=\displaystyle0
\end{array}\right.
\end{equation}
with
\begin{equation}\label{equ3.37}
\left\{\begin{array}{l}
\displaystyle\hat{M}_{I_{l}}=\displaystyle\frac{1}{c^{2}}\int d^{3}x'\hat{X'}_{I_{l}}\Big(T^{00}(\boldsymbol{x}')+T^{aa}(\boldsymbol{x}')\Big),\\
\displaystyle\hat{S}_{I_{l}}=\displaystyle\frac{1}{c}\int d^{3}x'\epsilon_{ab\langle i_{1}}\hat{X'}_{i_{2}\cdots i_{l}\rangle a}T^{0b}(\boldsymbol{x}'),\quad l\geq1
\end{array}\right.
\end{equation}
as the mass and spin multipole moments, where $\epsilon_{iab}$ is the totally antisymmetric Levi-Civita symbol and $\hat{X'}_{I_{l}}$ is the STF part of $X'_{I_{l}}:=x'_{i_{1}}x'_{i_{2}}\cdots x'_{i_{l}}$. By following the method in  Ref.~\cite{Wu:2017vvm}, the solutions to Eqs.~(\ref{equ3.34}) and (\ref{equ3.35}) for a spatially compact source are, respectively,
\begin{eqnarray}
\label{equ3.38}X^{(1)}(\boldsymbol{x})
&=&-\frac{m^{2}_{1}\kappa}{4\pi}\sum_{l=0}^{\infty}\frac{(-1)^l}{l!}\hat{Q}_{I_{l}}\hat\partial_{I_{l}}\left(\frac{\text{e}^{-m_{1}r}}{r}\right),\\
\label{equ3.39}P^{\mu\nu}(\boldsymbol{x})
&=&\frac{m^{2}_{2}\kappa}{4\pi}\sum_{l=0}^{\infty}\frac{(-1)^l}{l!}F_{\langle I_{l}\rangle}^{\mu\nu}\hat\partial_{I_{l}}\left(\frac{\text{e}^{-m_{2}r}}{r}\right)
\end{eqnarray}
with
\begin{eqnarray}
\label{equ3.40}\hat{Q}_{I_{l}}&=&\int \hat{X'}_{I_{l}}\delta_{l}(m_{1}r')T(\boldsymbol{x}')d^{3}x',\\
\label{equ3.41}F_{\langle I_{l}\rangle }^{\mu\nu}&=&\int \hat{X'}_{I_{l}}\delta_{l}(m_{2}r')S^{\mu\nu}(\boldsymbol{x}')d^{3}x'.
\end{eqnarray}
Here~\cite{Arfken1985},
\begin{eqnarray}
\label{equ3.42}&&\delta_{l}(z):=(2l+1)!!\bigg(\frac{d}{zdz}\bigg)^{l}\bigg(\frac{\sinh{z}}{z}\bigg)
\end{eqnarray}
satisfies
\begin{eqnarray}
%\label{equ3.43}&&\lim_{z\rightarrow0}\bigg(\frac{d}{zdz}\bigg)^{l}\bigg(\frac{\sinh{z}}{z}\bigg)=\frac{1}{(2l+1)!!},\\
\label{equ3.43}&&\lim_{z\rightarrow0}\delta_{l}(z)=1.
\end{eqnarray}

In order to acquire the gravitational field amplitude $h^{\mu\nu}$, one needs to resort to Eq.~(\ref{equ3.14}), and by further
employing Eqs.~(\ref{equ3.20}), (\ref{equ3.24}), (\ref{equ3.25}), and (\ref{equ3.31}), $h^{\mu\nu}$ can be expressed in terms of
$\tilde{h}^{\mu\nu}$, $P^{\mu\nu}$, and $X^{(1)}$, namely
\begin{eqnarray}
\label{equ3.44}h^{\mu\nu}=\tilde{h}^{\mu\nu}+\frac{2}{m_{2}^2}P^{\mu\nu}+\frac{2}{3m_{1}^2m_{2}^2}\partial^{\mu}\partial^{\nu}X^{(1)}-\frac{1}{3}\left(\frac{1}{m_{1}^2}+\frac{1}{m_{2}^2}\right)\eta^{\mu\nu}X^{(1)}.
\end{eqnarray}
Then, inserting Eqs.~(\ref{equ3.38}) and (\ref{equ3.39}),
\begin{eqnarray}
\label{equ3.45}h^{\mu\nu}=\tilde{h}^{\mu\nu}+\frac{4G}{c^4}\sum_{l=0}^{\infty}\frac{(-1)^l}{l!}\bigg[\hat{Q}_{I_{l}}\left(\frac{m_{1}^{2}+m_{2}^{2}}{6m_{2}^{2}}\eta^{\mu\nu}-\frac{1}{3m_{2}^{2}}\partial^{\mu}\partial^{\nu}\right)\hat\partial_{I_{l}}\left(\frac{\text{e}^{-m_{1}r}}{r}\right)+F_{\langle I_{l}\rangle }^{\mu\nu}\hat\partial_{I_{l}}\left(\frac{\text{e}^{-m_{2}r}}{r}\right)\bigg].
\end{eqnarray}
As previously noted, $\tilde{h}^{\mu\nu}$ is the counterpart of $h^{\mu\nu}$ in GR, and the second term is the correction to $\tilde{h}^{\mu\nu}$ in $F(X,Y,Z)$ gravity. Remember that the dimensions of the coefficients $F_{2},F_{3},$ and $F_{11}$ are all $[X]^{-1}$, so the above correction to $\tilde{h}^{\mu\nu}$ shows a Yukawa-like dependence on two characteristic lengths $m_{1}^{-1}$ and $m_{2}^{-1}$.
On the basis of Eqs.~(\ref{equ3.24}) and (\ref{equ3.25}), when $F(X,Y,Z)\rightarrow f(R)$, there are $m_{1}^{2}\rightarrow 1/(3F_{11})$ and $m_{2}\rightarrow+\infty$, and then, Eq.~(\ref{equ3.45}) recovers the corresponding result in $f(R)$ gravity~\cite{Wu:2017vvm}. As to the metric for the gravitational field, it is given by~\cite{Wu:2021uws}
\begin{eqnarray}
\label{equ3.46}g_{\mu\nu}=\eta_{\mu\nu}-\overline{h}_{\mu\nu}+o(h^{\mu\nu})
\end{eqnarray}
with
\begin{eqnarray}
\label{equ3.47}\overline{h}_{\mu\nu}:=h_{\mu\nu}-\frac{1}{2}\eta_{\mu\nu}h.
\end{eqnarray}
Thus, the metric for the gravitational field outside a spatially compact stationary source in $F(X,Y,Z)$ gravity is
\begin{equation}\label{equ3.48}
\left\{\begin{array}{ll}
\displaystyle g_{00}(\boldsymbol{x})&=\displaystyle -1+\frac{2}{c^{2}}U(\boldsymbol{x})-\frac{4}{c^{2}}V(\boldsymbol{x})+o(h^{\mu\nu}),\smallskip\\
\displaystyle g_{0i}(\boldsymbol{x})&=\displaystyle -\frac{4}{c^{3}}U^{i}(\boldsymbol{x})+\frac{4}{c^{3}}V^{i}(\boldsymbol{x})+o(h^{\mu\nu}),\smallskip\\
\displaystyle g_{ij}(\boldsymbol{x})&=\displaystyle \delta_{ij}\left(1+\frac{2}{c^{2}}U(\boldsymbol{x})\right)+\frac{4}{c^{2}}V^{ij}(\boldsymbol{x})+o(h^{\mu\nu}),
\end{array}\right.
\end{equation}
where the potentials $U(\boldsymbol{x})$, $V(\boldsymbol{x})$, $U^{i}(\boldsymbol{x})$, $V^{i}(\boldsymbol{x})$, and $V^{ij}(\boldsymbol{x})$ are, respectively, defined as
\begin{equation}\label{equ3.49}
\left\{\begin{array}{ll}
\displaystyle U(\boldsymbol{x})&:=\displaystyle G\sum_{l=0}^{\infty}\frac{(-1)^{l}}{l!}\hat{M}_{I_{l}}\partial_{I_{l}}\left( \frac{1}{r}\right),\smallskip\\
\displaystyle V(\boldsymbol{x})&:=\displaystyle \frac{G}{c^2}\sum_{l=0}^{\infty}\frac{(-1)^l}{l!}\bigg(\frac{1}{6}\hat{Q}_{I_{l}}\hat\partial_{I_{l}}\left(\frac{\text{e}^{-m_{1}r}}{r}\right)+F_{\langle I_{l}\rangle}^{00}\hat\partial_{I_{l}}\left(\frac{\text{e}^{-m_{2}r}}{r}\right)\bigg),\smallskip\\
\displaystyle U^{i}(\boldsymbol{x})&:=\displaystyle G\sum_{l=1}^{\infty}\frac{(-1)^{l}l}{(l+1)!}\epsilon_{iab}\hat{S}_{aI_{l-1}}\partial_{bI_{l-1}}\left( \frac{1}{r}\right),\smallskip\\
\displaystyle V^{i}(\boldsymbol{x})&:=\displaystyle \frac{G}{c}\sum_{l=0}^{\infty}\frac{(-1)^l}{l!}F_{\langle I_{l}\rangle}^{0i}\hat\partial_{I_{l}}\left(\frac{\text{e}^{-m_{2}r}}{r}\right),\smallskip\\
\displaystyle V^{ij}(\boldsymbol{x})&:=\displaystyle \frac{G}{c^2}\sum_{l=0}^{\infty}\frac{(-1)^l}{l!}\bigg(\hat{Q}_{I_{l}}\left(\frac{1}{6}\delta_{ij}+\frac{1}{3m_{2}^{2}}\partial_{i}\partial_{j}\right)\hat\partial_{I_{l}}\left(\frac{\text{e}^{-m_{1}r}}{r}\right)-F_{\langle I_{l}\rangle}^{ij}\hat\partial_{I_{l}}\left(\frac{\text{e}^{-m_{2}r}}{r}\right)\bigg)
\end{array}\right.
\end{equation}
with
\begin{equation}\label{equ3.50}
\left\{\begin{array}{ll}
\displaystyle F_{\langle I_{l}\rangle}^{00}&=\displaystyle \int \hat{X'}_{I_{l}}\delta_{l}(m_{2}r')\left(T^{00}(\boldsymbol{x}')+\frac{1}{3}T(\boldsymbol{x}')\right)d^{3}x',\\
\displaystyle F_{\langle I_{l}\rangle}^{0i}&=\displaystyle \int \hat{X'}_{I_{l}}\delta_{l}(m_{2}r')T^{0i}(\boldsymbol{x}')d^{3}x',\\
\displaystyle F_{\langle I_{l}\rangle}^{ij}&=\displaystyle \int \hat{X'}_{I_{l}}\delta_{l}(m_{2}r')\left(T^{ij}(\boldsymbol{x}')-\frac{1}{3}\delta_{ij}T(\boldsymbol{x}')+\frac{1}{3m_{2}^2}\frac{\partial^2T(\boldsymbol{x}')}{\partial x'_{i}\partial x'_{j}}\right)d^{3}x'.
\end{array}\right.
\end{equation}
Equations~(\ref{equ3.37}), (\ref{equ3.40}), and (\ref{equ3.50}) indicate that integrations in the results of the present paper are performed only over the domain occupied by the source, which is differently from those obtained with the Green function method in Refs.~\cite{stabile2015,Capozziello:2009ss,Stabile:2010mz}. Such feature could lead to a wider application of the result in this paper.

From Eqs.~(\ref{equ3.24}) and (\ref{equ3.25}), we observe that when $F(X,Y,Z)$ gravity reduces to GR, both $m_{1}$ and $m_{2}$ tend to positive infinity, which means that the potentials $V(\boldsymbol{x})$, $V^{i}(\boldsymbol{x})$, and $V^{ij}(\boldsymbol{x})$ vanish, and thus,
\begin{equation}\label{equ3.51}
\left\{\begin{array}{ll}
\displaystyle g_{00}^{GR}(\boldsymbol{x})&=\displaystyle -1+\frac{2}{c^{2}}U(\boldsymbol{x})+o(h^{\mu\nu}),\smallskip\\
\displaystyle g_{0i}^{GR}(\boldsymbol{x})&=\displaystyle -\frac{4}{c^{3}}U^{i}(\boldsymbol{x})+o(h^{\mu\nu}),\smallskip\\
\displaystyle g_{ij}^{GR}(\boldsymbol{x})&=\displaystyle \delta_{ij}\left(1+\frac{2}{c^{2}}U(\boldsymbol{x})\right)+o(h^{\mu\nu})
\end{array}\right.
\end{equation}
is exactly the metric in GR. Therefore, in the expression~(\ref{equ3.48}), the terms, not related to the potentials $V(\boldsymbol{x})$, $V^{i}(\boldsymbol{x})$, and $V^{ij}(\boldsymbol{x})$, constitute the GR-like part, and the remaining terms, being the correction to the GR-like part in $F(X,Y,Z)$ gravity, constitute the modified part. Obviously, the modified part is characterized by the two characteristic lengths $m_{1}^{-1}$ and $m_{2}^{-1}$, which are dependent on the value of derivatives of $F$ with respect to curvature invariants.
The multipole expansion of the metric potential in $F(X,Y,Z)$ gravity for a spatially compact stationary source can be directly read off from Eqs.~(\ref{equ3.48}) and (\ref{equ3.49}), namely,
\begin{equation}\label{equ3.52}
\Phi(\boldsymbol{x}):=U(\boldsymbol{x})-2V(\boldsymbol{x})=G\sum_{l=0}^{\infty}\frac{(-1)^l}{l!}\hat{M}_{I_{l}}\hat{\partial}_{I_{l}}\left( \frac{1}{r}\right)-G\sum_{l=0}^{\infty}\frac{(-1)^l}{l!}\bigg[\frac{1}{3c^2}\hat{Q}_{I_{l}}\hat\partial_{I_{l}}\left(\frac{\text{e}^{-m_{1}r}}{r}\right)+\frac{2}{c^2}F_{\langle I_{l}\rangle}^{00}\hat\partial_{I_{l}}\left(\frac{\text{e}^{-m_{2}r}}{r}\right)\bigg],
\end{equation}
where on the right-hand side, the first term, associated with the mass multipole moments $\hat{M}_{I_{l}}$, is identical to the metric potential in GR, and the remaining term represents the Yukawa-like corrections of $F(X,Y,Z)$ gravity to GR. Obviously, two additional sets of mass-type source multipole moments $\hat{Q}_{I_{l}}$ and $F_{\langle I_{l}\rangle}^{00}$ appear in the Yukawa-like corrections, and  Eqs.~(\ref{equ3.40}) and (\ref{equ3.50}) show that in the expressions of these multipole moments, the integrations are always modulated by a common radial factor $\delta_{l}(z)$ related to the source distribution. When the gravitational field is generated by a ball-like source and the distance from the center-of-mass of the source is much less than the scale associated with the volume occupied by the source, we are allowed to ignore all terms generated by the dipole moment and its higher-order analogues. Thus, there is
\begin{equation}\label{equ3.53}
\Phi(\boldsymbol{x})=G\bigg( \frac{\hat{M}}{r}-\frac{1}{3c^2}\frac{\hat{Q}\text{e}^{-m_{1}r}}{r}-\frac{2}{c^2}\frac{F^{00}\text{e}^{-m_{2}r}}{r}\bigg)
\end{equation}
with
\begin{eqnarray}
\label{equ3.54}\hat{M}&=&\frac{1}{c^{2}}\int d^{3}x'\Big(T^{00}(\boldsymbol{x}')+T^{aa}(\boldsymbol{x}')\Big),\\
\label{equ3.55}\hat{Q}&=&\int\frac{\sinh{(m_{1}r')}}{m_{1}r'}T(\boldsymbol{x}')d^{3}x',\\
\label{equ3.56}F^{00}&=&\int \frac{\sinh{(m_{2}r')}}{m_{2}r'}S^{00}(\boldsymbol{x}')d^{3}x'.
\end{eqnarray}
Obviously, even at the monopole order in $F(X,Y,Z)$ gravity, the effects of the distribution of matter within the source is significant, which is completely different from the case in GR. Hence, it is clear that the metric~(\ref{equ3.48}) can indicate the influence of the size and shape of a realistic source on the external gravitational field.
%Since the metric obtained in this section is applicable for the external gravitational field of any spatially compact stationary source, it is certain that our result will make $F(X,Y,Z)$ gravity have a wider range of applications.
%In fact, the similar conclusion has been reported in Refs.~\cite{stabile2015,Stabile:2010mz} by means of the metric at Newtonian order for point-like source. Now, if a point-like source with mass $M$ is situated at the origin, and to Newtonian order, we have $T^{00}(\boldsymbol{x})=Mc^2\delta^{3}(\boldsymbol{x}),\ T^{0i}(\boldsymbol{x})=0$, and $T^{ij}(\boldsymbol{x})=0$, where  $\delta^{3}(\boldsymbol{x})$ is the three-dimensional Dirac delta function. Plugging them into the expressions of $U(\boldsymbol{x})$ and $V(\boldsymbol{x})$, one could obtain the metric potential in $F(X,Y,Z)$ gravity for point-like source,
%\begin{equation}\label{equ3.491}
%U(\boldsymbol{x})-2V(\boldsymbol{x})=GM\left(\frac{1}{r}+\frac{1}{3}\frac{\text{e}^{-m_{1}r}}{r}-\frac{4}{3}\frac{\text{e}^{-m_{2}r}}{r}\right),
%\end{equation}
%which is identical to that in Refs.~\cite{stabile2015,Stabile:2010mz}. Although the result for point-like source is very meaningful in analyzing the structural characteristics of the metric potential in $F(X,Y,Z)$ gravity, it is not suitable to be used to explore phenomena happening in the gravitational field outside a realistic gravitating source because the effect of the size and shape of the source is extremely crucial in general.

As previously mentioned, $F(X,Y,Z)$ gravity, being a generic fourth-order theory of gravity, contains a large number of sub-models, such as GR, $f(R)$ gravity, and $f(R,\mathcal{G})$ gravity, etc., and as a consequence, the metrics for the external gravitational field of a spatially compact stationary source in these models could be directly obtained from Eq.~(\ref{equ3.48}) under certain conditions. By this means, the metric~(\ref{equ3.51}) in GR has already been derived, and it is observed that both characteristic lengths $m_{1}^{-1}$ and $m_{2}^{-1}$ disappear in GR. As to $f(R)$ gravity, when $F(X,Y,Z)\rightarrow f(R)$, the power series~(\ref{equ3.5}) reduces to
\begin{equation}
\label{equ3.57}f(R)=R+\frac{1}{2}f_{11}R^2+\cdots,
\end{equation}
and then, from Eqs.~(\ref{equ3.24}) and (\ref{equ3.25}), there are $m_{1}^{2}\rightarrow 1/(3f_{11})$ and $m_{2}\rightarrow+\infty$. Thus, inserting them into Eqs.~(\ref{equ3.48}) and (\ref{equ3.49}), one can deduce the metric in $f(R)$ gravity,
\begin{equation}\label{equ3.58}
\left\{\begin{array}{ll}
\displaystyle g_{00}^{f(R)}(\boldsymbol{x})&=\displaystyle -1+\frac{2}{c^{2}}U(\boldsymbol{x})-\frac{4}{c^{2}}V_{f(R)}(\boldsymbol{x})+o(h^{\mu\nu}),\smallskip\\
\displaystyle g_{0i}^{f(R)}(\boldsymbol{x})&=\displaystyle -\frac{4}{c^{3}}U^{i}(\boldsymbol{x})+o(h^{\mu\nu}),\smallskip\\
\displaystyle g_{ij}^{f(R)}(\boldsymbol{x})&=\displaystyle \delta_{ij}\left(1+\frac{2}{c^{2}}U(\boldsymbol{x})+\frac{4}{c^{2}}V_{f(R)}(\boldsymbol{x})\right)+o(h^{\mu\nu})
\end{array}\right.
\end{equation}
with
\begin{equation}\label{equ3.59}
\left\{\begin{array}{ll}
\displaystyle V_{f(R)}(\boldsymbol{x})&:=\displaystyle \frac{G}{6c^2}\sum_{l=0}^{\infty}\frac{(-1)^l}{l!}\hat{Q}_{I_{l}}\hat\partial_{I_{l}}\left(\frac{\text{e}^{-m_{1}r}}{r}\right),\smallskip\\
\displaystyle V^{i}_{f(R)}(\boldsymbol{x})&:=\displaystyle 0,\bigskip\\
\displaystyle V^{ij}_{f(R)}(\boldsymbol{x})&:=\displaystyle \delta_{ij}V_{f(R)}(\boldsymbol{x}),
\end{array}\right.
\end{equation}
which is exactly the same as that in Ref.~\cite{Wu:2017vvm}. These expressions imply that since the characteristic length $m_{2}^{-1}$ vanishes, there is only one massive propagation with mass $m_{1}$ in $f(R)$ gravity. Now, let us examine $f(R,\mathcal{G})$ gravity.
We postulate that
\begin{equation}\label{equ3.60}
F(X,Y,Z)=f(R,\mathcal{G}):=R+f_{2}\mathcal{G}+\frac{1}{2}\left(f_{11}R^{2}+2f_{12}R\mathcal{G}+f_{22}\mathcal{G}^2\right)+\cdots,
\end{equation}
and due to $\mathcal{G}=X^2-4Y+Z$, there are
\begin{equation}\label{equ3.61}
\left\{\begin{array}{ll}
\displaystyle F_{11}&=\displaystyle f_{11}+2f_{2},\smallskip\\
\displaystyle F_{2}&=\displaystyle -4f_{2},\smallskip\\
\displaystyle F_{3}&=\displaystyle f_{2}.
\end{array}\right.
\end{equation}
Substituting these expressions in Eqs.~(\ref{equ3.24}) and (\ref{equ3.25}), we acquire that $m_{1}^{2}=1/(3f_{11})$ and $m_{2}=+\infty$, which explicitly indicates the metric in $f(R,\mathcal{G})$ gravity is also Eq.~(\ref{equ3.58}). Therefore, it is concluded that
the metric in $f(R,\mathcal{G})$ gravity is the same as that in $f(R)$ gravity, which is consistent with the fact that the Gauss-Bonnet scalar $\mathcal{G}$, being a topological invariant, has no contribution to the gravitational field dynamics.

Equations~(\ref{equ3.48}) and (\ref{equ3.51}) explicitly indicate that the potential functions $V(\boldsymbol{x})$, $V^{i}(\boldsymbol{x})$, and $V^{ij}(\boldsymbol{x})$ are the corrections to $g_{00}^{GR}(\boldsymbol{x})$, $g_{0i}^{GR}(\boldsymbol{x})$, and $g_{ij}^{GR}(\boldsymbol{x})$, respectively, and hence, from them, the additional gravitational degrees of freedom appearing in $F(X,Y,Z)$ gravity could be analyzed. In practice, by applying the metric~(\ref{equ3.48}) to some specific phenomenon, one can directly explore the effects of these new gravitational degrees of freedom. For example, for a gyroscope moving around the source in geodesic motion, one is able to utilize the metric~(\ref{equ3.48}) to derive its spin's angular velocity of precession. In GR, the precessional angular velocity of the gyroscope spin is decomposed in the geodetic and Schiff contributions~\cite{poisson2014gravity}. According to the conventional method in Ref.~\cite{MTW1973}, when the precessional angular velocity of the gyroscope spin is derived in $F(X,Y,Z)$ gravity, it could be expected that the geodetic precessional angular velocity would be corrected by the potential functions $V(\boldsymbol{x})$ and $V^{ij}(\boldsymbol{x})$, whereas the Schiff precessional angular velocity would be corrected by the potential function $V^{i}(\boldsymbol{x})$. As a result, by comparing these results with the data of the gyroscopic experiment, e.g., GP-B, the constraints on the coefficients of curvature invariants in the gravitational Lagrangian of $F(X,Y,Z)$ gravity can be derived, and then, substituting these constraints back in the theoretical results, one will obtain the effects of the new gravitational degrees of freedom in $F(X,Y,Z)$ gravity. Similarly, the metric~(\ref{equ3.48}) can also be applied to the anomalous perihelion advance of Mercury, the gravitational redshift of light, and the light bending, etc. By comparing the theoretical results with the experimental or observational data, the further effects of the new gravitational degrees of freedom in $F(X,Y,Z)$ gravity will also be derived.
\black
%In the future, when one studies some specific phenomenon within the framework of $F(X,Y,Z)$ gravity, the increment or reduction of the characteristic lengths in its various sub-models will have a great influence on final theoretical results. Thus, by comparing these theoretical results with the experimental or observational data, the constraints on the coefficients of curvature invariants in the gravitational Lagrangians of these models may be obtained, and in this manner, a large class of fourth-order theories of gravity could be assessed in detail.
%%%%%%%%%%%%%%%%%%%%%%%%%%%%%%%%%%%%%%%%%%%%%%%%%%%%%%%
%%%%%%%%%%%%%%%%%%%%%%%%%%%%%%%%%%%%%%%%%%%%%%%%%%%%%%%
%%% The third chapter                              %%%%
%%%%%%%%%%%%%%%%%%%%%%%%%%%%%%%%%%%%%%%%%%%%%%%%%%%%%%%
%%%%%%%%%%%%%%%%%%%%%%%%%%%%%%%%%%%%%%%%%%%%%%%%%%%%%%%
\section{Conclusions~\label{Sec:fifth}}
In this paper, $F(X,Y,Z)$ gravity, a generic fourth-order theory of gravity involving curvature invariants $X=R,\ Y=R_{\mu\nu}R^{\mu\nu},$ and $Z=R_{\mu\nu\rho\sigma}R^{\mu\nu\rho\sigma}$, has been considered. Conducting a study on $F(X,Y,Z)$ gravity is of great importance because it provides a general theoretical framework for many famous gravitational models, such as GR, Starobinsky gravity, $f(R)$ gravity, and $f(R,\mathcal{G})$ gravity, etc. In order to deal with various phenomena within $F(X,Y,Z)$ gravity, the first thing we need to do is to deduce the metric for the external gravitational field of a gravitating source.
In Refs.~\cite{stabile2015,Capozziello:2009ss,Stabile:2010mz}, some versions of the metric are provided with the Green function method under the weak-field and slow-motion approximation, but since integrations involved in these results usually need to be carried out over the whole space, they seem to be inconvenient to apply in practice. In this paper, this problem is rehandled by applying the symmetric and trace-free formalism in terms of the irreducible Cartesian tensors under the weak-field approximation, and the metric for the external gravitational field of a spatially compact stationary source is obtained.

By applying the weak-field approximation, the linearized gravitational field equations of $F(X,Y,Z)$ gravity are gained firstly, and how to further deal with them and find their solutions is one of the main task of the present paper. Following the method in Refs.~\cite{Wu:2017vvm,Wu:2018hjx,Wu:2018jve,Wu:2021uws}, we propose a new type of gauge condition in $F(X,Y,Z)$ gravity, and after imposing it, the gravitational field amplitude $h^{\mu\nu}$, associated with the metric $g_{\mu\nu}$, is successfully expressed in terms of the effective gravitational field amplitude $\tilde{h}^{\mu\nu}$, the tensor $P^{\mu\nu}$ relevant to the linearized Ricci tensor $R^{\mu\nu(1)}$, and the linearized Ricci scalar $X^{(1)}$, where $\tilde{h}^{\mu\nu}$ satisfies d'Alembert equation and both $P^{\mu\nu}$ and $X^{(1)}$ satisfy Klein-Gordon equations with external sources. By invoking the stationary solutions to d'Alembert equation and Klein-Gordon equation in Ref.~\cite{Wu:2017vvm}, the metric for the gravitational field outside a spatially compact stationary source in $F(X,Y,Z)$ gravity is derived.
%尽管本文的主题是给出静态解，但是线性方程的这种处理方式也提供了在F引力中讨论引力辐射的基础。
Compared with those in Refs.~\cite{stabile2015,Capozziello:2009ss,Stabile:2010mz}, integrations involved in the result of the present paper are performed only over the volume occupied by the source distribution in the application process, which could lead to a wider application of the result in this paper. It should be pointed out that when one tries to discuss the gravitational radiation within $F(X,Y,Z)$ gravity, such simplification of the linearized gravitational field equations by imposing the new type of gauge condition is also valid, which implies that in applications of $F(X,Y,Z)$ gravity, this type of treatment under the weak-field approximation is extremely important.

Since the effective gravitational field amplitude $\tilde{h}^{\mu\nu}$, satisfying the wave equation, behaves just as its counterpart in GR, the terms related to $\tilde{h}^{\mu\nu}$ in the expression of the metric constitute the GR-like part, and the remaining terms constitute the modified part, where the latter is the correction to the former in $F(X,Y,Z)$ gravity. It is shown that the metric is characterized by two characteristic lengths depending on the value of derivatives of $F$ with respect to curvature invariants, which means that in $F(X,Y,Z)$ gravity, there are two massive propagations in general.
From the expression of the metric, the multipole expansion of the metric potential in $F(X,Y,Z)$ gravity for a spatially compact stationary
source can be directly read off. In this expansion, the GR-like part, identical to the metric potential in GR, is associated with the mass multipole moments, and the modified part,  representing the Yukawa-like corrections to the GR-like part, is associated with two additional sets of mass-type source multipole moments. The expressions of these two sets of source multipole moments display that the integrations are always modulated by a common radial factor related to the source, which implies that differently from the case in GR, even at the monopole order in $F(X,Y,Z)$ gravity, the effects of the distribution of matter within the source is significant.
%Moreover, since this metric is applicable for the external gravitational field of any spatially compact stationary source, it is suitable to be used to explore phenomena happening in the gravitational field outside a realistic gravitating source because the effect of the size and shape of the source is extremely crucial. As a consequence, it is certain that our result will make $F(X,Y,Z)$ gravity have a wider range of applications.

Finally, in GR, $f(R)$ gravity, and $f(R,\mathcal{G})$ gravity, the metrics for the external gravitational field of a spatially compact stationary source are derived from the one in $F(X,Y,Z)$ gravity because these theories are all the sub-models of $F(X,Y,Z)$ gravity. When $F(X,Y,Z)$ gravity reduces to $f(R)$ or $f(R,\mathcal{G})$ gravity, it is proved that one characteristic length of the metric in $F(X,Y,Z)$ gravity disappears, and the metric in $f(R,\mathcal{G})$ gravity is identical to that in $f(R)$ gravity, which confirms the fact that
that the Gauss-Bonnet scalar $\mathcal{G}$, as a topological invariant, has no contribution to the gravitational field dynamics.
When $F(X,Y,Z)$ gravity further reduces to GR, both the characteristic lengths of the metric disappear, and the GR-like part is exactly the result in GR.
It is straightforward to explore the effects of the additional gravitational degrees of freedom appearing in $F(X,Y,Z)$ gravity by applying the metric to some specific phenomenon. Such a typical example is that for a gyroscope moving around the source in geodesic motion, one is able to utilize the metric to derive its spin's angular velocity of precession. According to the conventional method in Ref.~\cite{MTW1973}, when the precessional angular velocity of the gyroscope spin is derived in $F(X,Y,Z)$ gravity, the precessional angular velocity in GR would be corrected by the potential functions in the modified part of the metric. By comparing these results with the data of the gyroscopic experiment, e.g., GP-B, the effects of the new gravitational degrees of freedom in $F(X,Y,Z)$ gravity will be obtained.
The metric can also be applied to the anomalous perihelion advance of Mercury, the gravitational redshift of light, and the light bending, etc, and in the same way, the further effects of the new gravitational degrees of freedom in $F(X,Y,Z)$ gravity will also be derived.
\black
%It could be expected that the increment or reduction of the characteristic lengths in these sub-models of $F(X,Y,Z)$ gravity will have a great impact on their theoretical predictions for a physical phenomenon.

As noted above, $F(X,Y,Z)$ gravity is a generic fourth-order theory of gravity, and it contains a large number of sub-models, and for these models, the metric obtained in the present paper is a universal outcome, which actually provides a powerful tool for us to analyze the gravitational phenomena happening in the gravitational field outside a realistic stationary source in these models. Although only the metric for stationary source is derived, it is sufficient to be employed to explain many phenomena, such as the light bending. One can also seek to find the external metric for non-stationary source under the weak-field and slow-motion approximation, but a basic fact is that such metric is only valid within the near zone of the source~\cite{eric2018}, so it seems that our present result will make $F(X,Y,Z)$ gravity have a wider range of applications.
In addition, based on the weak-field approximation developed in this paper, gravitational waves could also be discussed within the framework of $F(X,Y,Z)$ gravity. It can be expected that compared with the result in GR, further gravitational modes will appear in $F(X,Y,Z)$ gravity. In the future, with the help of the data from gravitational wave observations, such as LISA experiment and Einstein Telescope, these new gravitational modes in $F(X,Y,Z)$ gravity would be detected~\cite{LISACosmologyWorkingGroup:2022wjo,Lombriser:2016yzn,Astashenok:2020qds} so that we are able to determine that these modes are retained or rejected.
\black
%if the metrics in various sub-models of $F(X,Y,Z)$ gravity are applied to dealing with some specific phenomenon, by comparing the resulting theoretical result with the experimental or observational data, the constraints on the coefficients of curvature invariants in the gravitational Lagrangians of these models may be obtained, and therefore, our result can also be used to assess in detail a large class of fourth-order theories of gravity.

%%%%%%%%%%%%%%%%%%%%%%%%%%%%%%%%%%%%%%%%%%%%%%%%%%%%%%%
%%%%%%%%%%%%%%%%%%%%%%%%%%%%%%%%%%%%%%%%%%%%%%%%%%%%%%%
%%% Conclusions and discussions                    %%%%
%%%%%%%%%%%%%%%%%%%%%%%%%%%%%%%%%%%%%%%%%%%%%%%%%%%%%%%
%%%%%%%%%%%%%%%%%%%%%%%%%%%%%%%%%%%%%%%%%%%%%%%%%%%%%%%
%\acknowledgments{}
\begin{acknowledgments}
This work was supported by the National Natural Science Foundation of China (Grants Nos.~12105039 and 12035016).
\end{acknowledgments}
%%%%%%%%%%%%%%%%%%%%%%%%%%%%%%%%%%%%%%%%%%%%%%%%%%%%%%%
%%%%%%%%%%%%%%%%%%%%%%%%%%%%%%%%%%%%%%%%%%%%%%%%%%%%%%%

\end{document}